\documentclass[a4paper,11pt]{article}
\pdfoutput=1
\usepackage{jheppub}
\usepackage{graphicx}
\usepackage{subfigure}
\usepackage{dcolumn}
\usepackage{verbatim}
\usepackage{amsmath}
\usepackage{amssymb}
\usepackage{float}

\title{\boldmath Simplest Neutrino Mixing from $S_4$ Symmetry}

\author[a]{R.~Krishnan,}
\author[a]{P.~F.~Harrison}
\author[b]{and W.~G.~Scott}

\affiliation[a]{University of Warwick,\\Coventry, CV4 7AL, UK}
\affiliation[b]{Rutherford Appleton Laboratory,\\Chilton, Didcot, Oxon, OX11 0QX, UK}

\emailAdd{k.rama@warwick.ac.uk}
\emailAdd{p.f.harrison@warwick.ac.uk}
\emailAdd{w.g.scott@rl.ac.uk}

\abstract{In 2004, two of us proposed a texture, the ``Simplest'' neutrino mass matrix, which predicted $\sin{\theta_{13}}=\sqrt{2\Delta m^2_{sol}/3\Delta m^2_{atm}}$ and $\delta_{CP}=90^\circ$. Using today's measured values for neutrino mass-squared differences, this prediction gives $\sin^2\! 2{\theta_{13}}\simeq 0.086^{+0.003}_{-0.006}$, compared with a measured value, found by averaging the results of the Daya Bay and RENO experiments, of $\sin^2\! 2{\theta_{13}}=0.093\pm0.010$. Here we present a specific model based on $S_4$ symmetry leading to this successful texture in the context of the type-1 see-saw mechanism, assuming Majorana neutrinos. In this case, slightly different predictions are obtained relating $\theta_{13}$ to the light neutrino masses, which are in accord with current experimental limits and testable at future experiments. Large $CP$ asymmetries remain a generic prediction of the texture.}

\begin{document}


\maketitle
\flushbottom

\section{Introduction}
\label{sec:model2intro}

Leptonic mixing is characterised by two large mixing angles, $\theta_{12}\simeq 35^{\circ}$ and $\theta_{23}\simeq 45^{\circ}$, and one small angle, $\theta_{13}$. For several years, the data on neutrino oscillations were compatible with $\theta_{13}=0$, and the data together were approximated by the tribimaximal (TBM) mixing matrix, proposed in 2002~\cite{TBM}. TBM has been used by many authors as a starting point for model building. The Daya Bay Reactor Neutrino Experiment~\cite{DayaBay} has recently measured the value for the reactor mixing angle, $\sin^22\theta_{13}=0.092\pm0.016\text{~(stat.)}\pm0.005\text{~(syst.)}$, showing definitively that $\theta_{13}$ is non-zero. Similarly, the RENO Experiment~\cite{RENO} made a compatible measurement, $\sin^2 2 \theta_{13} = 0.113 \pm 0.013\text{~(stat.)} \pm 0.019\text{~(syst.)}$. T2K~\cite{T2K}, Double Chooz~\cite{DCHOOZ} and MINOS~\cite{MINOS} experiments also measured consistent non-zero values for $\theta_{13}$. In models of leptonic mixing based on discrete symmetries starting with TBM, it is possible to  generate non-zero $\theta_{13}$ by introducing higher order corrections, but in the generic case, the deviations produced should be of the same order for all three mixing angles~\cite{altarelli}. However, since the experimentally allowed deviation of $\theta_{12}$ from its TBM value, $\sin^2 \theta_{12}=1/3$, is small, it is difficult to generate the relatively large experimental value of $\theta_{13}$ in this way. 

Anticipating an eventual non-zero value for $\theta_{13}$, two of us proposed several generalisations of the TBM texture in 2002~\cite{TXM}, in which the condition $\theta_{13}=0$ was relaxed in various ways. For example, in a basis in which the charged lepton mass matrix is diagonal, a hermitian neutrino mass matrix which has $\mu$-$\tau$ reflection symmetry~\cite{mutau} and democracy (each of its rows and columns sums to a common value~\cite{demo}) leads to ``tri$\chi$maximal'' mixing (T$\chi$M)~\cite{TXM,Xing},
\begin{equation}
\left|U_{\text{T}\chi\text{M}}\right|=\left(\begin{matrix}|\sqrt{\frac{2}{3}}\cos \chi| & |\frac{1}{\sqrt{3}}| & |\sqrt{\frac{2}{3}}\sin \chi|\\
|-\frac{\cos \chi}{\sqrt{6}}\mp i\frac{\sin \chi}{\sqrt{2}}| & |\frac{1}{\sqrt{3}}| & |\pm i\frac{\cos \chi}{\sqrt{2}}-\frac{\sin \chi}{\sqrt{6}}|\\
|-\frac{\cos \chi}{\sqrt{6}}\pm i\frac{\sin \chi}{\sqrt{2}}| & |\frac{1}{\sqrt{3}}| & |\mp i\frac{\cos \chi}{\sqrt{2}}-\frac{\sin \chi}{\sqrt{6}}|
\end{matrix}\right).
\label{eq:txmform}
\end{equation}
The only free parameter in T$\chi$M mixing is the angle $\chi$. From Eq.~(\ref{eq:txmform}) it is straightforward to obtain the standard PDG mixing angles in terms of the parameter $\chi$,
\begin{align}
&|U_{e3}^\dagger|^2=\sin^2 \theta_{13}=\frac{2}{3}\sin^2 \chi \label{eq:theta13ask1}\\
&|U_{e2}^\dagger|^2=\sin^2 \theta_{12} \cos^2 \theta_{13}=\frac{1}{3}\\
&|U_{\mu3}^\dagger|^2=\sin^2 \theta_{23} \cos^2 \theta_{13}=\frac{\sin^2 \chi}{6}+\frac{\cos^2 \chi}{2} \implies \sin^2 \theta_{23}=\frac{1}{2}\\
&\delta_{CP}=\pm\frac{\pi}{2}
\end{align}

The most general hermitian mass matrix having $\mu$-$\tau$ reflection symmetry and democracy is given by:
\begin{equation}\label{eq:generalhermitian}
M_H=
a\left(\begin{matrix}1 & \pm ik & \mp ik\\
       \mp ik& 0 & 1\pm ik\\
       \pm ik & 1\mp ik & 0
\end{matrix}\right)+b\left(\begin{matrix}0 & 1 & 1\\
       1 & 0 & 1\\
       1 & 1 & 0
\end{matrix} \right) + c\left(\begin{matrix}1 & 0 & 0\\
       0 & 1 & 0\\
       0 & 0 & 1
\end{matrix}\right)
\end{equation}
where $k$, $a$, $b$, and $c$ are real parameters. In other words the unitary matrix which diagonalises $M_H$, Eq.~(\ref{eq:generalhermitian}), will have the \text{T}$\chi$\text{M} form, Eq.~(\ref{eq:txmform}).

The mass matrix arising from the Majorana mass term for the neutrinos should be complex-symmetric. Therefore we wish to determine the general complex-symmetric mass matrix that generates T$\chi$M. One way to achieve this is to multiply the matrix $M_H$ by the $\mu$-$\tau$ exchange operator $P$ to get the complex-symmetric matrix $M_S$:
\begin{equation}\label{eq:generalsymmetric}
M_H \, P= M_S=
a\left(\begin{matrix}1 & \mp ik & \pm ik\\
       \mp ik & 1\pm ik & 0\\
        \pm ik & 0 & 1\mp ik
\end{matrix}\right)+b\left(\begin{matrix}0 & 1 & 1\\
       1 & 1 & 0\\
       1 & 0 & 1
\end{matrix} \right) + c\left(\begin{matrix}1 & 0 & 0\\
       0 & 0 & 1\\
       0 & 1 & 0
\end{matrix}\right)
\end{equation}
where $P=\left(\begin{smallmatrix}1 & 0 & 0\\
       0 & 0 & 1\\
       0 & 1 & 0
\end{smallmatrix}\right)$.

The only free parameter in T$\chi$M, $\chi$, is uniquely determined by the real parameter $k$ in the matrix $M_S$, Eq.~(\ref{eq:generalsymmetric}). It can be shown that
\begin{equation}\label{eq:theta13ask2}
\cos 2\chi = \frac{1}{\sqrt{1+3k^2}}.
\end{equation}
Using Eqs.~(\ref{eq:theta13ask1},~\ref{eq:theta13ask2}) we get
\begin{equation}\label{eq:theta13ask}
\sin^2\!{\theta_{13}}=\frac{1}{3}\left(1-\frac{1}{\sqrt{1+3k^2}}\right)
\end{equation}

As with any complex-symmetric matrix, $M_S$ can be diagonalised using a unitary matrix and its transpose to give real positive eigenvalues:
\begin{equation}\label{eq:generalhermitiandiag}
U^{\dag} \, M_S \, U^*=\,\text{Diag}(|a\sqrt{1+3k^2}-b+c|,|a+2b+c|,|-a\sqrt{1+3k^2}-b+c|)
\end{equation}
where $U$ has the T$\chi$M form, i.e.~$|U|=|U_{\text{T}\chi\text{M}}|$. The unitary matrix $U$ which diagonalises the matrix $M_S$ also diagonalises each of its three terms independently. The first term of $M_S$ alone is in fact, sufficient to generate T$\chi$M, having three distinct eigenvalues. The second and the third terms of $M_S$ give two degenerate and three degenerate eigenvalues respectively.

A special case of the T$\chi$M texture, known as ``Simplest'' neutrino mixing was proposed in 2004~\cite{simplest} (after having been introduced and discussed briefly already in 2002~\cite{TXM}) by setting $b=0$ in the texture of Eq.~(\ref{eq:generalhermitian}). Its eigenvalues are given by the RHS of Eq.~(\ref{eq:generalhermitiandiag}) with $b=0$. Simplest neutrino mixing yields an exact and very straightforward relation between the reactor mixing angle (see Eq.~(\ref{eq:theta13ask})) and the eigenvalues, $e_i$:
\begin{equation}
\sin^2\!{\theta_{13}}=\frac{2}{3}\frac{(e_2-e_1)}{(e_3-e_1)}.
\label{eq:simplestPred1}
\end{equation}
In the original publications \cite{TXM, simplest}, this texture was proposed for $M_\nu^2:=M_\nu M_\nu^\dag$, in which case
the eigenvalues are the neutrino masses-squared, resulting in the very successful prediction:
\begin{equation}
\sin{\theta_{13}}=\sqrt{\frac{2}{3}\frac{\Delta m^2_{\text{sol}}}{\Delta m^2_{\text{atm}}}}.
\label{eq:simplestPred2}
\end{equation}
\begin{eqnarray}
\text{i.e.}~\sin^2\! 2{\theta_{13}}&=& 0.086^{+0.003}_{-0.006}\qquad{(\text{Predicted in 2002/2004~\cite{TXM, simplest}})}\\
~\text{cf.}~\sin^2\! 2{\theta_{13}}&=&0.093\pm0.010\quad{(\text{Measured in 2012~\cite{DayaBay, RENO}}}).
\end{eqnarray}

In view of the very encouraging phenomenological success of the ``Simplest'' mixing hypothesis, we here propose a model for it based on the symmetric group of degree four ($S_4$). In the model, however, we make the following two changes with respect to the original hypothesis: i)~the ``Simplest'' mass matrix form is adopted for the mass matrix itself (as opposed to its hermitian square); ii)~in order to exploit the type-I see-saw mechanism a Majorana mass term is assumed (coupling between two heavy right-handed neutrinos). Thus, we get a Majorana neutrino mass matrix of the following ``Simplest'' complex-symmetric form:
\begin{equation}\label{eq:complexsymm}
M_\nu(\text{Majorana})=a\left(\begin{matrix}1 & \mp ik & \pm ik\\
       \mp ik & 1\pm ik & 0\\
        \pm ik & 0 & 1\mp ik
\end{matrix}\right)+ c\left(\begin{matrix}1 & 0 & 0\\
       0 & 0 & 1\\
       0 & 1 & 0
\end{matrix}\right).
\end{equation}
In the following section we construct such a Majorana mass matrix assuming symmetry under the $S_4$ group. The neutrino Dirac mass matrix (coupling between the left-handed and the right-handed neutrinos) is assumed to be proportional to the identity. We show that this model has a phenomenology compatible with experiment, and we use it to predict the masses of the light neutrinos.

\section{The group $S_4$ and the $\mu$-$\tau$ rotated basis}
\label{sec:model2intro}
\begin{figure}[H]
\centering
\mbox{
\subfigure[]{
   \includegraphics[scale =0.75] {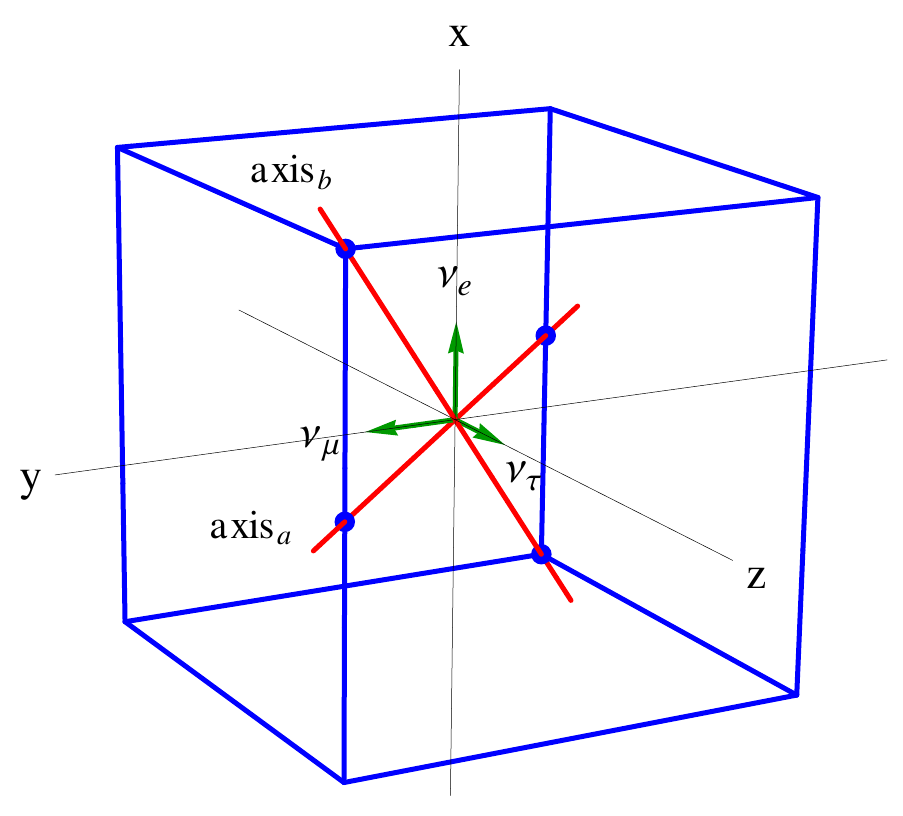}
   \label{fig:S4a}
 }
\quad
 \subfigure[]{
   \includegraphics[scale =0.74] {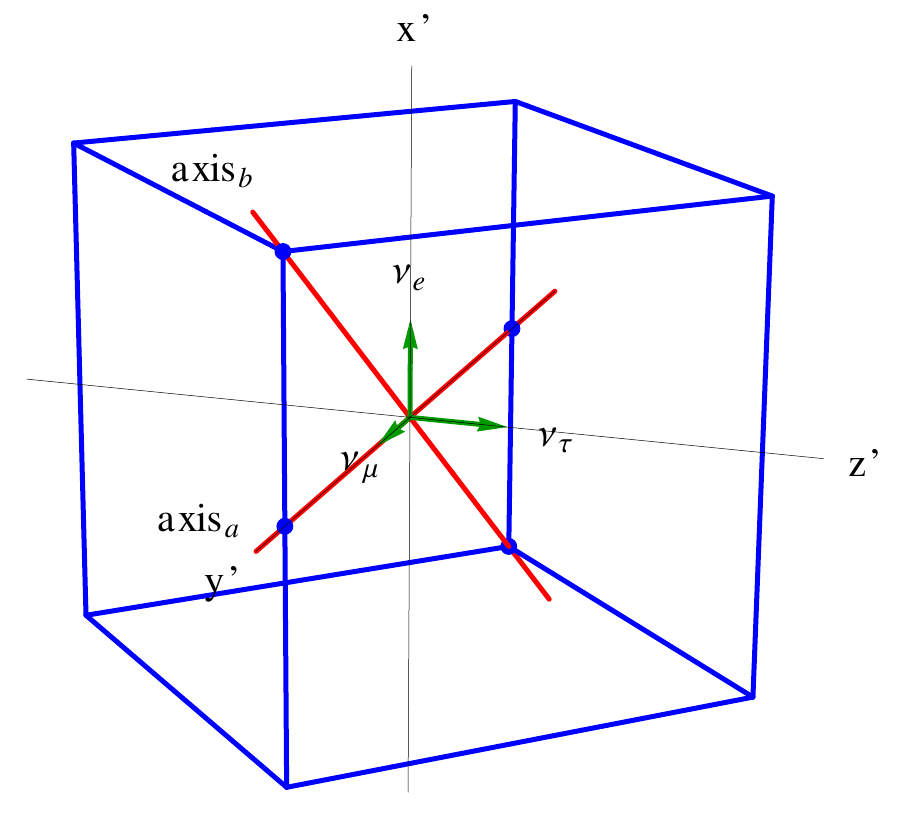}
   \label{fig:S4b}
 }
}
\caption{Octahedral symmetry in the $\mu$-$\tau$ rotated basis $(x',y',z')$.}\label{fig:S4}
\end{figure}

$S_4$, the group of permutations of four objects, is the symmetry group of the cube and the octahedron. In an abstract form, the group has the presentation~\cite{book}
\begin{equation} 
\langle \boldsymbol{a},\boldsymbol{b} | \boldsymbol{a}^2 = \boldsymbol{b}^3 = (\boldsymbol{a}\boldsymbol{b})^4 = e \rangle,
\end{equation}
where $\boldsymbol{a}$, $\boldsymbol{b}$ and $\boldsymbol{a}\boldsymbol{b}$ represent the orientation-preserving rotations of a cube through angles $\pi$, $\frac{2\pi}{3}$ and $\frac{\pi}{2}$ respectively. This is shown in Fig.~\ref{fig:S4}, where $\text{axis}_a$ and $\text{axis}_b$ correspond to the generators $\boldsymbol{a}$ and $\boldsymbol{b}$ respectively. A detailed study of the $S_4$ group can be found in~\cite{S4Table}. The $S_4$ group has been used extensively as the flavour symmetry group in model building, e.g.~\cite{S41,S42,S43,S44,S45,S46}.

We may work in a basis with:
\begin{equation}\label{eq:nonrotgen}
\boldsymbol{a} =
\left(\begin{matrix}-1 & 0 & 0\\
       0 & 0 & 1\\
       0 & 1 & 0
\end{matrix}\right)\\, \quad \boldsymbol{b}=
\left(\begin{matrix}0 & 1 & 0\\
       0 & 0 & 1\\
       1 & 0 & 0
\end{matrix}\right),
\end{equation}
as is used frequently in flavour models. In this basis the coordinate system is oriented such that the
coordinate axes are normal to the faces of the cube, as shown in Fig.~\ref{fig:S4a}. So the $x$, $y$ and $z$ coordinate axes are the symmetry axes of $\frac{\pi}{2}$-rotations. In a model constructed with a neutrino triplet $(\nu_e,\nu_\mu,\nu_\tau)$ defined parallel to the coordinate axes $(x, y, z)$  in the above basis, $\nu_e$, $\nu_\mu$ and $\nu_\tau$ are simply the invariant eigenstates (eigenstates with eigenvalue equal to $+1$) of these particular $\frac{\pi}{2}$-rotations \footnote{Abstractly they are also the eigenvectors of the corresponding elements of the three-element conjugacy class $C_3$ ($\pi$-rotations about axes passing through face centres).}. This choice of eigenstates is straightforward, and is the one used in most models using this group so far. It is however, by no means the only choice, and there is no reason why we should not define the flavour basis states in a different way. To construct a model for deriving the ``Simplest'' texture, it will prove useful to define the $\nu_\mu$ and $\nu_\tau$ basis states rotated by an angle $\frac{\pi}{4}$ relative to the $x$, $y$ and $z$ coordinate axes defined above, using the rotation matrix,
\begin{equation}\label{eq:R}
R=\left(\begin{matrix}1 & 0 & 0\\
       0 & \frac{1}{\sqrt{2}} & \frac{1}{\sqrt{2}}\\
       0 & -\frac{1}{\sqrt{2}} & \frac{1}{\sqrt{2}}
\end{matrix}\right).
\end{equation}
The matrix $R$ represents a rotation about the $x$ axis by an angle $\pi/4$, relative to the cube. We also rotate the $y$ and $z$ coordinate axes to align with the new $\nu_\mu$ and $\nu_\tau$ flavour basis states respectively. In the rotated coordinate system, $(x^\prime,y^\prime,z^\prime)$ in Fig.~\ref{fig:S4b}, we have the group generators:
\begin{equation}\label{eq:rotgen}
\boldsymbol{a}\rightarrow R.\boldsymbol{a}.R^\dagger =\left(\begin{matrix}-1 & 0 & 0\\
       0 & 1 & 0\\
       0 & 0 & -1
\end{matrix}\right), \quad \boldsymbol{b}\rightarrow R.\boldsymbol{b}.R^\dagger=\left(\begin{matrix}0 & \frac{1}{\sqrt{2}} & -\frac{1}{\sqrt{2}}\\
       \frac{1}{\sqrt{2}} & \frac{1}{2} & \frac{1}{2}\\
       \frac{1}{\sqrt{2}} & -\frac{1}{2} & -\frac{1}{2}
\end{matrix}\right). 
\end{equation}
The state $\nu_e$ is unchanged and still corresponds to the $\frac{\pi}{2}$-rotation symmetry of the cube about the $x$-axis. However, $\nu_\mu$ and $\nu_\tau$ are no longer invariant eigenstates of $\frac{\pi}{2}$-rotations, but rather are invariant under $\pi$ rotations of the cube, as may be seen in Fig.~\ref{fig:S4b}. We call this new basis, the $\mu$-$\tau$ rotated basis.

The three-dimensional representation of $S_4$ corresponding to the rotational
symmetries of the cube is denoted by $\boldsymbol{3'}$\cite{S4Table}. Thus the neutrino triplet $(\nu_e,\nu_\mu,\nu_\tau)$ belongs to the $\boldsymbol{3'}$ representation. A Majorana mass term contains two neutrino fields, and thus it is of interest to consider the tensor product decomposition of two $\boldsymbol{3'}$s. This decomposition is as follows:
\begin{equation}\label{eq:expansionexpr}
\boldsymbol{3'}\times\boldsymbol{3'}=\boldsymbol{1}+\boldsymbol{2}+\boldsymbol{3}+\boldsymbol{3'},
\end{equation}
where $\boldsymbol{1}$ is the trivial representation, and in the $\mu$-$\tau$ rotated basis we have:
\begin{align}\label{eq:expansion}
\chi_1&=\frac{1}{\sqrt{3}}(\nu_e.\nu_e+\nu_\mu.\nu_\mu+\nu_\tau.\nu_\tau), \quad &\chi_2&=\left(\begin{matrix}-\sqrt{\frac{2}{3}}\nu_e.\nu_e+\frac{1}{\sqrt{6}}\nu_\mu.\nu_\mu+\frac{1}{\sqrt{6}}\nu_\tau.\nu_\tau\\
       \frac{1}{\sqrt{2}}(\nu_\mu.\nu_\tau+\nu_\tau.\nu_\mu )
\end{matrix}\right)\notag \\ 
\chi_3&=\left(\begin{matrix}\frac{1}{\sqrt{2}}(\nu_\mu.\nu_\mu-\nu_\tau.\nu_\tau )\\
       \frac{1}{\sqrt{2}}(\nu_e.\nu_\mu+\nu_\mu.\nu_e )\\
       \frac{1}{\sqrt{2}}(\nu_\tau.\nu_e+\nu_e.\nu_\tau )\\
\end{matrix}\right),\quad&\chi_3'&=\left(\begin{matrix}\frac{1}{\sqrt{2}}(\nu_\mu.\nu_\tau-\nu_\tau.\nu_\mu )\\
       \frac{1}{\sqrt{2}}(\nu_\tau.\nu_e-\nu_e.\nu_\tau )\\
       \frac{1}{\sqrt{2}}(\nu_e.\nu_\mu-\nu_\mu.\nu_e )\\ 
\end{matrix}\right)=0,\notag\\
\end{align}
where the bi-linears $\chi_1$, $\chi_2$, $\chi_3$, $\chi_3'$ transform as $\boldsymbol{1}$, $\boldsymbol{2}$, $\boldsymbol{3}$, $\boldsymbol{3'}$ respectively. The product $\nu_i.\nu_j$ is the Lorentz invariant product of the right-handed neutrino Weyl spinors. Obviously the terms in $\chi_3'$ in Eq.~(\ref{eq:expansion}) vanish. The Kronecker products of all the irreducible representations (irreps) of $S_4$ and all the relevant Clebsch-Gordan coefficients are given in Appendix~A.

We now assume three types of flavons, $\phi_1$, $\phi_2$ and $\phi_3$ which transform as $\boldsymbol{1}$, $\boldsymbol{2}$ and $\boldsymbol{3}$ respectively, allowing us to write an invariant mass term:
\begin{equation}\label{eq:trichiInv}
\text{Inv}=c_1 \chi_1 \phi_1 + c_2 \chi_2^T \phi_2 + c_3 \chi_3^T \phi_3,
\end{equation}
where $c_1$, $c_2$ and $c_3$ are constants. Once the flavons acquire specific forms of vacuum expectation values (VEVs), the required mass matrix can be obtained from the invariant mass term given in Eq.~(\ref{eq:trichiInv}). Suppose the flavons get VEVs 
\begin{equation}\label{eq:flavvev}
\langle\phi_1\rangle=1, \langle\phi_2\rangle=\left(-\frac{1}{2},\frac{\sqrt{3}}{2}\right), \langle\phi_3\rangle=\left(1,-1,1\right),
\end{equation}
the mass matrix obtained will be 
\begin{equation}\label{eq:majmassform}
M=c_1 I + \frac{c_2\sqrt{3}}{2\sqrt{2}} \left(\begin{matrix}\frac{2}{3} & 0 & 0\\
       0 & -\frac{1}{3} & 1\\
       0 & 1 & -\frac{1}{3}
\end{matrix}\right) + \frac{c_3}{\sqrt{2}} \left(\begin{matrix}0 & -1 & 1\\
       -1 & 1 & 0\\
       1 & 0 & -1
\end{matrix}\right),
\end{equation}
which is in the same form as Eq.~(\ref{eq:complexsymm}), if $c_1$ and $c_2$ are real and $c_3$ is imaginary (unlike in Eq.~(\ref{eq:complexsymm}), here the part proportional to the identity and the parts which are traceless are given separately.). The more general case where $c_1$, $c_2$ and $c_3$ are arbitrary complex numbers leads to ``$S_3$~group mixing''~\cite{demo}. Note that the constraint of $c_1$ and $c_2$ being real and $c_3$ being imaginary is an assumption made in order to obtain T$\chi$M which is a special case of $S_3$~group mixing where $CP$ violation is maximal. In the model there is no group theoretical justification for such an assumption since $c_1$, $c_2$ and $c_3$ are simply free parameters. Of course, embedding the flavons $\phi_1$, $\phi_2$ and $\phi_3$ in a single representation of a larger group and then breaking the group into $S_4$ giving the three separate flavons would lead to contraints among $c_1$, $c_2$ and $c_3$, but such considerations are beyond the scope of this paper. The details of how to obtain the flavon VEVs, Eq.~(\ref{eq:flavvev}), using the minimisation of flavon potentials are given in Appendix~B.

\section{The model}

A number of models and parameterisations have been proposed, e.g.~\cite{M0,M1,M2,M3,M4,M5,M6,M7,M8}, to accommodate the non-zero $\theta_{13}$. In this paper we construct the model in a Standard Model framework with the addition of heavy right-handed neutrinos. Through a type-1 seesaw mechanism, light Majorana neutrinos are produced. The fermion and flavon content of the model with representations to which they belong is given in the Table~\ref{tab:flavourcontent}. We construct a diagonal mass matrix for the charged leptons using the mass term given by Eq.~(\ref{eq:mclept}). The $C_2$ flavour symmetries $C_{2e}$, $C_{2\mu}$ and $C_{2\tau}$ ensure that the flavons $\phi_{3e}'$, $\phi_{3\mu}'$ and $\phi_{3\tau}'$ couple to only $e_R$, $\mu_R$ and $\tau_R$ respectively.  In order to construct the Dirac mass term containing the right-handed neutrinos and the left-handed lepton doublets, we postulate a singlet flavon $\phi_1^-$. We also introduce another $C_2$ symmetry $C_{2D}$ to allow only the singlet flavon $\phi_1^-$ to enter the Dirac mass term, Eq.~(\ref{eq:mdl2dirac}), and thus to make the neutrino Dirac mass matrix proportional to the identity. The Majorana mass term containing the right-handed neutrinos, Eq.~(\ref{eq:mnurnur}), leads to a mass matrix of the form given by Eq.~(\ref{eq:majmassform}) as explained in the previous section. The Standard Model Higgs field is assigned to the trivial representation of $S_4$.

For the charged leptons, the mass term is of the form
\begin{equation}\label{eq:mclept}
\left(y_e L^\dagger e_R \phi_{3e}'+y_\mu L^\dagger \mu_R \phi_{3\mu}'+y_\tau L^\dagger \tau_R \phi_{3\tau}'\right)\frac{H}{\Lambda}+H.C.
\end{equation}
where $H$ is the standard model Higgs, $\Lambda$ is the cut-off scale and the $y_i$ are coupling constants. The VEVs of the Higgs and the flavons break the weak gauge symmetry and the flavour symmetry respectively. With the Higgs getting the VEV, $(0, h_o)$, and the flavons, $\phi_{3e}'$, $\phi_{3\mu}'$ and $\phi_{3\tau}'$, getting the VEVs proportional to $(1,0,0)$, $(0,1,0)$ and $(0,0,1)$ respectively, we obtain the required masses $m_e$, $m_\mu$ and $m_\tau$ for the charged leptons. In Appendix~B we provide for illustrative purposes an example of flavon potentials which on minimisation give rise to the above mentioned VEVs.

\begin{table}[H]
\begin{center}
\begin{tabular}{|c|c|c|c|c|c|c|c|c|c|c|c|c|}
\hline
	&$e_R$	&$\mu_R$&$\tau_R$&$L$	&$\nu_R$&$\phi_1^-$&$\phi_1$&$\phi_2$&$\phi_3$&$\phi_{3e}'$&$\phi_{3\mu}'$&$\phi_{3\tau}'$\\
\hline
$S_4$	&$\boldsymbol{1}$&$\boldsymbol{1}$&$\boldsymbol{1}$&$\boldsymbol{3'}$&$\boldsymbol{3'}$&$\boldsymbol{1}$&$\boldsymbol{1}$&$\boldsymbol{2}$&$\boldsymbol{3}$&$\boldsymbol{3'}$&$\boldsymbol{3'}$&$\boldsymbol{3'}$\\
\hline
$C_{2e}$	&$-1$	&$1$	&$1$	&$1$	&$1$	&$1$	&$1$	&$1$		&$1$	&$-1$	&$1$	&$1$	\\
\hline
$C_{2\mu}$	&$1$	&$-1$	&$1$	&$1$	&$1$	&$1$	&$1$	&$1$		&$1$	&$1$	&$-1$	&$1$	\\
\hline
$C_{2\tau}$	&$1$	&$1$	&$-1$	&$1$	&$1$	&$1$	&$1$	&$1$		&$1$	&$1$	&$1$	&$-1$	\\
\hline
$C_{2D}$	&$-1$	&$-1$	&$-1$	&$-1$	&$1$	&$-1$	&$1$	&$1$		&$1$	&$1$	&$1$	&$1$	\\
\hline
\end{tabular}
\end{center}
\caption{The flavour structure of the model. $L$ are the three left-handed lepton weak isospin doublets and $\nu_R$ are the three right-handed heavy neutrinos. The $\boldsymbol{3'}$ representations are in the $\mu$-$\tau$ rotated basis and the $\boldsymbol{2}$ and the $\boldsymbol{3}$ representations are in the basis given by the tensor product expansion in Eqs.~(\ref{eq:expansionexpr},\,\ref{eq:expansion}).}
\label{tab:flavourcontent}
\end{table}

The Dirac mass term for the neutrinos takes the form
\begin{equation}\label{eq:mdl2dirac}
y_w L^\dagger \nu_R \frac{\phi_1^-}{\Lambda} \tilde{H}+H.C.
\end{equation}
where $\tilde{H}$ is the conjugate Higgs and $y_w$ is a coupling.
We also have the Majorana mass term for the neutrinos:
\begin{equation}\label{eq:mnurnur}
\left(y_1 \chi_1 \phi_1 + y_2 \chi_2^T \phi_2 + i y_3 \chi_3^T \phi_3\right)\frac{1}{\Lambda},
\end{equation}
where the $\chi_i$ are given by the expressions in Eqs.~(\ref{eq:expansion}) and the $y_i$ ($i=1$, 2, 3) are couplings leading to very heavy right-handed Majorana masses. The flavons $\phi_1$, $\phi_2$ and $\phi_3$ getting the VEVs given in Eqs.~(\ref{eq:flavvev}) and also having $\langle \phi_1^-\rangle=1$ result in the following $6\times6$ mass matrix $M$ for the neutrinos:
\begin{equation}
\nu_\alpha^T \, M \, \nu^\alpha
\end{equation}

\begin{equation}
M=\left(\begin{matrix}0 & M_\text{Dir}\\
       M_\text{Dir} & M_\text{Maj}
\end{matrix}\right), \quad \nu=\left(\begin{matrix}\nu_{L}^*\\
	\nu_{R}
\end{matrix}\right)
\end{equation}
with $\nu_L=(\nu_e,\nu_\mu,\nu_\tau)$ the left-handed neutrino flavour eigenstates where $M_\text{Dir} = \frac{y_w}{2\Lambda} I$ and $M_\text{Maj}$ is of the required form given in Eqs.~(\ref{eq:complexsymm}) and (\ref{eq:majmassform}).
Here, $M_\text{Maj}$ is at a very high mass scale and $M_\text{Dir}$ is of order the weak scale, so that the seesaw mechanism comes into play. It can be shown \cite{Majorana} that the effective mass matrix, $M_\text{ss}$, generated via the seesaw mechanism for the left-handed light neutrinos is in the form
\begin{equation}
M_\text{ss}=-M_\text{Dir}M^{-1}_\text{Maj}M_\text{Dir}.
\end{equation}
The matrix $iU^*$ diagonalises $M_\text{ss}$ giving light neutrino masses proportional to $\frac{1}{e_1}$, $\frac{1}{e_2}$, $\frac{1}{e_3}$ where $e_1$, $e_2$ and $e_3$ are the eigenvalues of $M_\text{Maj}$.

It is to be noted that we may rotate the fermion and the flavon fields from the $\mu$-$\tau$ rotated basis to the conventional basis and obtain the same physical result, even though in the conventional basis the charged-lepton mass matrix will not be diagonal and the Majorana mass term for the neutrinos will not be in the ``Simplest'' texture. Therefore the use of $\mu$-$\tau$ rotated basis in the model ensures that the charged-lepton mass matrix is diagonal and the Majorana mass term is in the ``Simplest'' texture. 

\section{Fitting the model with experimental data}
\begin{figure}[H]
\begin{center}
\includegraphics[scale=0.75]{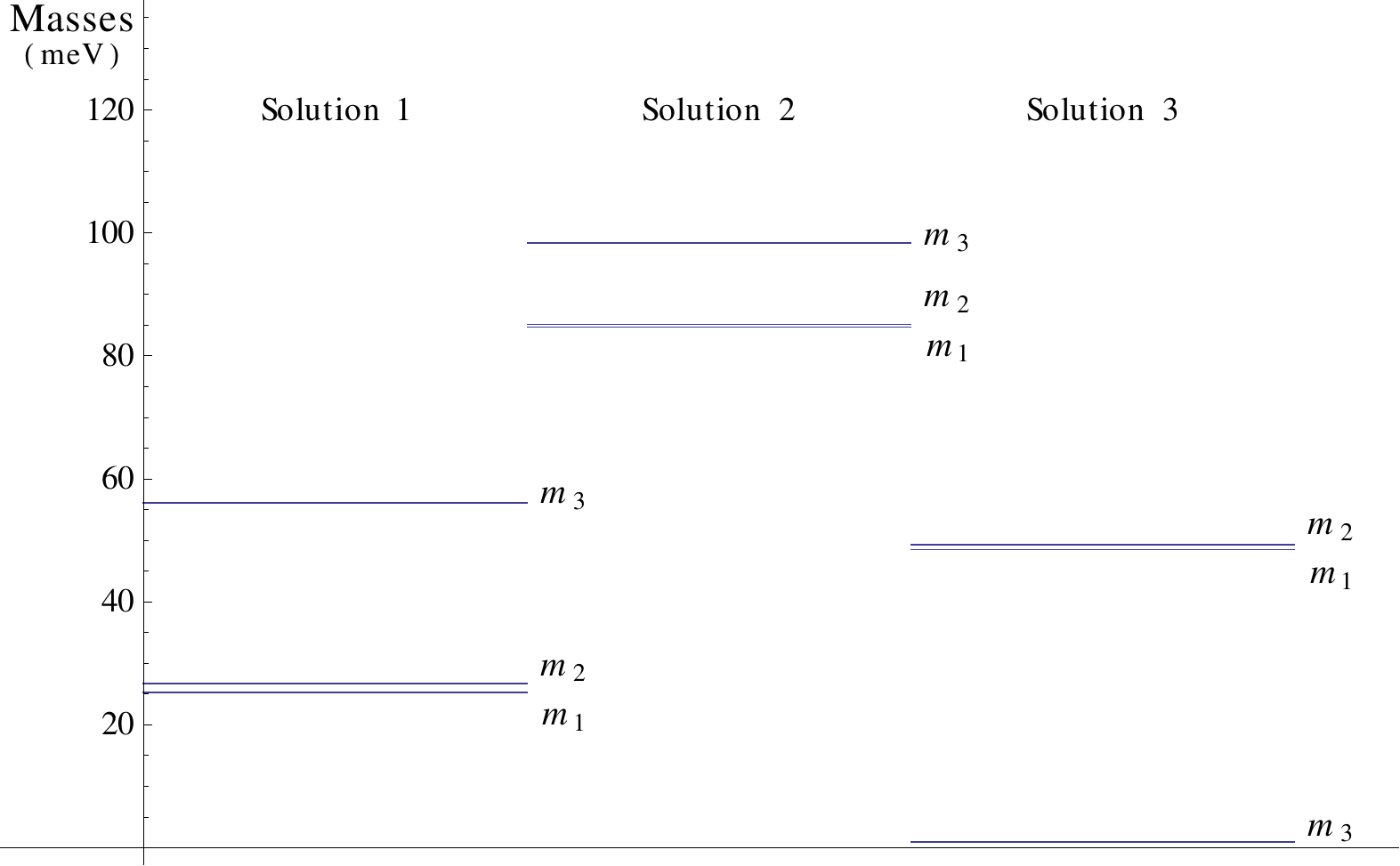}
\caption{The predicted values of the neutrino masses corresponding to the best fit $\quad \quad \quad \quad \quad(m_2^2-m_1^2= 75.9$~meV$^2$, $|m_3^2-m_2^2|= 2430$ meV$^2$, $\sin^22\theta_{13}=0.098)$. Case~1 ($r= 0.4101$) and case~2 ($r=14.452$) are in normal hierarchy. Case~3 ($r=-1.0405$) is in inverted hierachy.}
\label{fig:masseshierarchy}
\end{center}
\end{figure}

The squared differences of the light neutrino masses are known experimentally, $m_{2}^2-m_{1}^2= 75.9\pm 2.1$~meV$^2$, $|m_{3}^2-m_{2}^2|= 2430\pm 130$~meV$^2$. The eigenvalues of the Majorana mass matrix, $M_\text{Maj}$ from Eq.~(\ref{eq:complexsymm}), are
\begin{equation}
e_1=c+a\sqrt{1+3 k^2}, \quad e_2=c+a, \quad e_3=c-a\sqrt{1+3 k^2}
\end{equation}
($a$, $c$ and $k$ are real in our model, given our earlier assumptions). Since the light neutrino masses are inversely proportional to these eigenvalues, we get
\begin{gather}\label{eq:msdr}
\frac{m_2^2-m_1^2}{m_3^2-m_2^2}=\pm\frac{75.9\pm 2.1}{2430\pm 130}=\frac{\frac{1}{e_2^2}-\frac{1}{e_1^2}}{\frac{1}{e_3^2}-\frac{1}{e_2^2}}=-\frac{(r-s)^2(1-s)(1+2r+s)}{(r+s)^2(1+s)(1+2r-s)}
\end{gather}
where $r=c/a$ and $s=\sqrt{1+3k^2}$. Using Eq.~(\ref{eq:theta13ask}) we can calculate the parameter $k$ given the reactor mixing angle $\theta_{13}$. Substituting the value of $k$ in Eq.~(\ref{eq:msdr}) and solving for the parameter $r$, we can predict the values of the light neutrino masses. Eq.~(\ref{eq:msdr}) is cubic in $r$ giving three separate real solutions for normal hierarchy and one for inverted hierarchy. One of the normal hierarchy solutions gives the wrong sign for the solar mass-squared difference, leaving three remaining solutions. Thus the light neutrino masses predicted by the model fall into three sets. These results are shown in Fig.~\ref{fig:masseshierarchy} where the best fit values are used. 

\begin{figure}[H]
\begin{center}
\includegraphics[scale=1.3]{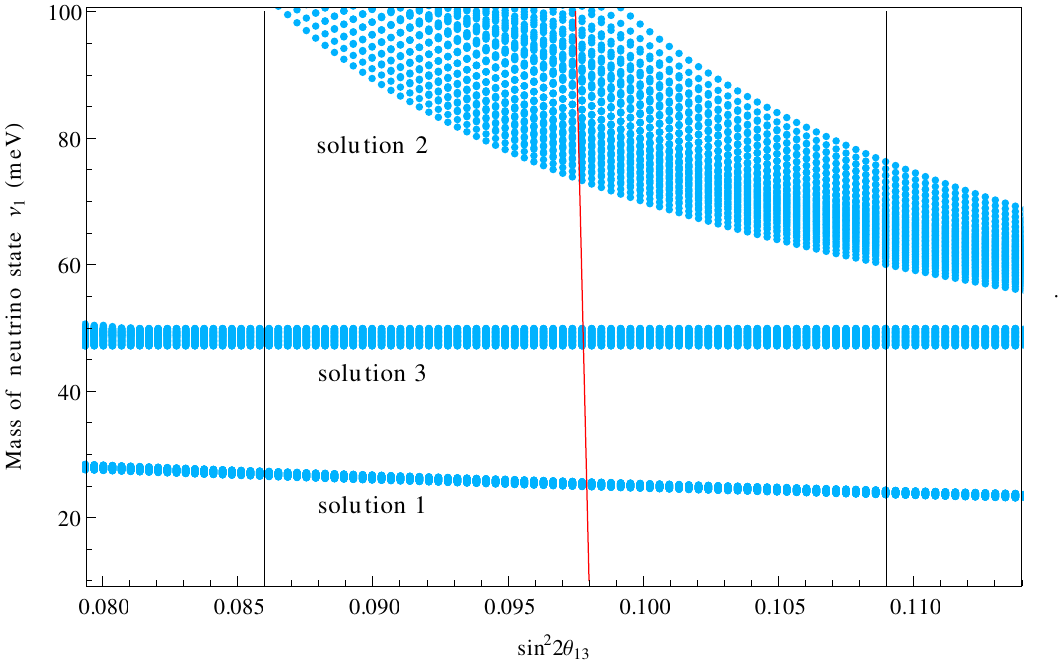}
\caption[The predicted mass of neutrino state $\nu_1$ along with experimental errors]{The predicted value of $m_1$ (the mass of the neutrino eigenstate $\nu_1$) vs the measured value of $\sin^22\theta_{13}$. The finite thickness of the bands is due to the errors in the measurement of the neutrino mass-squared differences. The red and the black lines indicate the best fit value and the errors on $\sin^22\theta_{13}$ respectively.}
\label{fig:model2fig}
\end{center}
\end{figure}

The error ranges of the mass of the neutrino eigenstate $\nu_1$ for the three solutions are shown in Fig.~\ref{fig:model2fig}. For solution~2 we do not consider a mass above $100~\text{meV}$ in order to keep our prediction compatible with the cosmological upper limit of the masses of the neutrinos. The WiggleZ Dark Energy Survey \cite{wigglez} gives the strongest cosmological limit so far, $\sum m_\nu<290~\text{meV}$. It should be emphasised that in any tri$\chi$maximal mixing (T$\chi$M) model, given the three mixing angles $\theta_{12}$, $\theta_{23}$, $\theta_{13}$, maximal $CP$ violation ($\delta_{CP}=\pm\frac{\pi}{2}$) is always guaranteed.

\section{Summary}
We use the ``Simplest'' texture of neutrino mass matrix to explain the recently measured non-zero reactor mixing angle. When used as a hermitian square of the mass matrix, this texture successfully predicts the observed reactor angle (with the help of the neutrino mass-squared differences). When used as a complex-symmetric Majorana mass term, the constraint among the masses and the mixing resulting from the texture can be used to predict the unknown lightest neutrino mass (with the help of the reactor angle). We get three solutions which are compatible with the measured mass-squared differences. In the model we exploit the ``$\mu$-$\tau$ rotated basis'', introduced here for the first time, to obtain the required texture for the mass matrix. Since the mixing is tri$\chi$maximal, the $CP$-violating phase is predicted to be $\pm\frac{\pi}{2}$. Large $CP$ violation such as this is potentially testable in future experiments. 

This work was supported by the UK Science and Technology Facilities Council (STFC). Two of us (PFH and RK) acknowledge the hospitality of the Centre for Fundamental Physics (CfFP) at the Rutherford Appleton Laboratory. RK acknowledges support from CfFP and the University of Warwick.

\section{Appendix A: Tensor product expansions of irreps of $S_4$}

The irreps of $S_4$ are $\boldsymbol{1}$, $\boldsymbol{2}$, $\boldsymbol{3}$, $\boldsymbol{3'}$. Eqs.~(\ref{eq:expansionexpr},~\ref{eq:expansion}) provide the tensor product expansion of $\boldsymbol{3'}\times\boldsymbol{3'}$. In Eq.~(\ref{eq:expansion}) we have used the $\mu$-$\tau$ rotated basis for $\boldsymbol{3'}$ as well as $\boldsymbol{3}$. For $\boldsymbol{3'}$, the $\mu$-$\tau$ rotated basis is defined using Eq.~(\ref{eq:rotgen}). For $\boldsymbol{3}$ we take
\begin{equation}
\boldsymbol{a} =
\left(\begin{matrix}1 & 0 & 0\\
       0 & 0 & -1\\
       0 & -1 & 0
\end{matrix}\right)\\, \quad \boldsymbol{b}=
\left(\begin{matrix}0 & 0 & 1\\
       1 & 0 & 0\\
       0 & 1 & 0
\end{matrix}\right),
\end{equation}
as the generators in the non-rotated basis and the transformation
\begin{equation}
\boldsymbol{a}\rightarrow R.\boldsymbol{a}.R^\dagger =\left(\begin{matrix}1 & 0 & 0\\
       0 & -1 & 0\\
       0 & 0 & 1
\end{matrix}\right), \quad \boldsymbol{b}\rightarrow R.\boldsymbol{b}.R^\dagger=\left(\begin{matrix}0 & \frac{1}{\sqrt{2}} & \frac{1}{\sqrt{2}}\\
       \frac{1}{\sqrt{2}} & \frac{1}{2} & -\frac{1}{2}\\
       -\frac{1}{\sqrt{2}} & \frac{1}{2} & -\frac{1}{2}
\end{matrix}\right)
\end{equation}
to define its $\mu$-$\tau$ rotated basis where $R$ is given in Eq.~(\ref{eq:R}). For $\boldsymbol{2}$ we have used a basis where
\begin{equation}\label{eq:2basis}
\boldsymbol{a} =
\left(\begin{matrix}1 & 0\\
       0 & -1
\end{matrix}\right)\\, \quad \boldsymbol{b}=
\left(\begin{matrix}-\frac{1}{2} & \frac{\sqrt{3}}{2}\\
       -\frac{\sqrt{3}}{2} & -\frac{1}{2}
\end{matrix}\right).
\end{equation}
Here we list all the non-trivial tensor product expansions of the various irreps expressed in the above mentioned bases, i.e.~$\boldsymbol{3'}$ and $\boldsymbol{3}$ in the $\mu$-$\tau$ rotated basis and $\boldsymbol{2}$ in the basis given in Eq.~(\ref{eq:2basis}).

\begin{subequations}
\begin{align}
\boldsymbol{3}\otimes\boldsymbol{3}&=\boldsymbol{1}\oplus\boldsymbol{2}\oplus\boldsymbol{3}\oplus\boldsymbol{3'}\\
\begin{split}
\left(\begin{matrix}a_1\\
       a_2\\
       a_3
\end{matrix}\right)\otimes \left(\begin{matrix}b_1\\
       b_2\\
       b_3
\end{matrix}\right)&=\frac{1}{\sqrt{3}}\left(a_1 b_1 + a_2 b_2+a_3 b_3\right)\oplus\left(\begin{matrix}\sqrt{\frac{2}{3}}a_1 b_1 -\frac{1}{\sqrt{6}} a_2 b_2-\frac{1}{\sqrt{6}}a_3 b_3\\
       \frac{1}{\sqrt{2}}a_2 b_3 +\frac{1}{\sqrt{2}} a_3 b_2
\end{matrix}\right)\\
&\quad\oplus\frac{1}{\sqrt{2}}\left(\begin{matrix}-a_2 b_2 + a_3 b_3\\
       -a_1 b_2 - a_2 b_1 \\
        a_1 b_3 + a_3 b_1
\end{matrix}\right)\oplus\frac{1}{\sqrt{2}}\left(\begin{matrix}-a_2 b_3 + a_3 b_2\\
        a_1 b_3 - a_3 b_1 \\
        a_1 b_2 - a_2 b_1
\end{matrix}\right)
\end{split}
\end{align}
\end{subequations}

\begin{subequations}
\begin{align}
\boldsymbol{3'}\otimes\boldsymbol{3}&=\boldsymbol{1'}\oplus\boldsymbol{2}\oplus\boldsymbol{3}\oplus\boldsymbol{3'}\\
\begin{split}
\left(\begin{matrix}a_1\\
       a_2\\
       a_3
\end{matrix}\right)\otimes \left(\begin{matrix}b_1\\
       b_2\\
       b_3
\end{matrix}\right)&=\frac{1}{\sqrt{3}}\left(a_1 b_1 + a_2 b_2-a_3 b_3\right)\oplus\left(\begin{matrix}-\frac{1}{\sqrt{2}}a_2 b_3 +\frac{1}{\sqrt{2}} a_3 b_2\\
\sqrt{\frac{2}{3}}a_1 b_1 -\frac{1}{\sqrt{6}} a_2 b_2+\frac{1}{\sqrt{6}}a_3 b_3
\end{matrix}\right)\\
&\quad\oplus\frac{1}{\sqrt{2}}\left(\begin{matrix}-a_2 b_3 - a_3 b_2\\
        a_1 b_3 + a_3 b_1 \\
        -a_1 b_2 + a_2 b_1
\end{matrix}\right)\oplus\frac{1}{\sqrt{2}}\left(\begin{matrix}a_2 b_2 + a_3 b_3\\
        a_1 b_2 + a_2 b_1 \\
        a_1 b_3 - a_3 b_1
\end{matrix}\right)
\end{split}
\end{align}
\end{subequations}

Note that for $\boldsymbol{1'}$ the generators are
\begin{equation}
\boldsymbol{a} = -1, \quad \boldsymbol{b}= 1.
\end{equation}

\begin{subequations}
\begin{align}
\boldsymbol{2}\otimes\boldsymbol{3'}&=\boldsymbol{3}\oplus\boldsymbol{3'}\\
\left(\begin{matrix}a_1\\
       a_2
\end{matrix}\right)\otimes \left(\begin{matrix}b_1\\
       b_2\\
       b_3
\end{matrix}\right)&=\left(\begin{matrix}-a_2 b_1\\
       -\frac{\sqrt{3}}{2}a_1 b_3 + \frac{1}{2} a_2 b_2\\
       \frac{\sqrt{3}}{2}a_1 b_2 - \frac{1}{2} a_2 b_3
\end{matrix}\right)\oplus\left(\begin{matrix}-a_1 b_1\\
       \frac{1}{2} a_1 b_2 + \frac{\sqrt{3}}{2} a_2 b_3\\
       \frac{1}{2} a_1 b_3+ \frac{\sqrt{3}}{2} a_2 b_2
\end{matrix}\right)\quad \quad \quad \quad \quad \quad
\end{align}
\end{subequations}

\begin{subequations}
\begin{align}
\boldsymbol{2}\otimes\boldsymbol{3}&=\boldsymbol{3}\oplus\boldsymbol{3'}\\
\left(\begin{matrix}a_1\\
       a_2
\end{matrix}\right)\otimes \left(\begin{matrix}b_1\\
       b_2\\
       b_3
\end{matrix}\right)&=\left(\begin{matrix}-a_1 b_1\\
       \frac{1}{2} a_1 b_2 -\frac{\sqrt{3}}{2} a_2 b_3\\
       \frac{1}{2} a_1 b_3 - \frac{\sqrt{3}}{2} a_2 b_2
\end{matrix}\right)\oplus\left(\begin{matrix}a_2 b_1\\
       -\frac{\sqrt{3}}{2} a_1 b_3 - \frac{1}{2} a_2 b_2\\
       \frac{\sqrt{3}}{2} a_1 b_2+ \frac{1}{2} a_2 b_3
\end{matrix}\right)\quad \quad \quad \quad \quad \quad
\end{align}
\end{subequations}

\begin{subequations}\label{eq:2times2}
\begin{align}
\boldsymbol{2}\otimes\boldsymbol{2}&=\boldsymbol{1}\oplus\boldsymbol{1'}\oplus\boldsymbol{2}\\
\left(\begin{matrix}a_1\\
       a_2
\end{matrix}\right)\otimes \left(\begin{matrix}b_1\\
       b_2
\end{matrix}\right)&=\frac{1}{\sqrt{2}}\left(a_1 b_1 + a_2 b_2\right)\oplus\frac{1}{\sqrt{2}}\left(a_1 b_2 - a_2 b_1\right)\oplus\frac{1}{\sqrt{2}}\left(\begin{matrix}-a_1 b_1 + a_2 b_2\\
       a_1 b_2 + a_2 b_1
\end{matrix}\right)
\end{align}
\end{subequations}

\begin{subequations}
\begin{align}
\boldsymbol{1'}\otimes\boldsymbol{3'}&=\boldsymbol{3}\\
a_1 \otimes \left(\begin{matrix}b_1\\
       b_2 \\
       b_3
\end{matrix}\right)&=\left(\begin{matrix}a_1 b_1\\
       a_1 b_2\\
       -a_1 b_3
\end{matrix}\right)\quad \quad \quad \quad \quad \quad \quad \quad \quad \quad \quad \quad \quad \quad \quad \quad \quad \quad \quad
\end{align}
\end{subequations}

\begin{subequations}
\begin{align}
\boldsymbol{1'}\otimes\boldsymbol{3}&=\boldsymbol{3'}\\
a_1 \otimes \left(\begin{matrix}b_1\\
       b_2 \\
       b_3
\end{matrix}\right)&=\left(\begin{matrix}a_1 b_1\\
       a_1 b_2\\
       -a_1 b_3
\end{matrix}\right)\quad \quad \quad \quad \quad \quad \quad \quad \quad \quad \quad \quad \quad \quad \quad \quad \quad \quad \quad
\end{align}
\end{subequations}

\begin{subequations}
\begin{align}
\boldsymbol{1'}\otimes\boldsymbol{2}&=\boldsymbol{2}\\
a_1 \otimes \left(\begin{matrix}b_1\\
       b_2
\end{matrix}\right)&=\left(\begin{matrix}-a_1 b_2\\
       a_1 b_1
\end{matrix}\right)\quad \quad \quad \quad \quad \quad \quad \quad \quad \quad \quad \quad \quad \quad \quad \quad \quad \quad \quad
\end{align}
\end{subequations}

\section{Appendix B: Flavon Potentials}

In this Appendix we show that we can obtain the required VEVs for the various flavons using the minimisation of the flavon potentials. Here we neither analyse all the possible invariant terms exhaustively nor provide the most general flavon potential. The example flavon potentials are provided for illustrative purpose only. 

Defining the flavon $\phi_2$ = $( \phi_{2}^1, \phi_{2}^2)$ and using Eq.~(\ref{eq:2times2}) we get a second degree doublet $\left(-(\phi_{2}^1)^2+(\phi_{2}^2)^2,2\phi_{2}^1 \phi_{2}^2\right)$. Combining it with the original doublet we construct a third degree invariant, $-(\phi_2^1)^3+3\phi_2^1 (\phi_2^2)^2$. Along with the term $(\phi_2^1)^2+(\phi_2^2)^2$ (which is basically $U(1)$ invariant), we may construct the potential
\begin{equation}
V(\phi_2)\propto\left((\phi_2^1)^2+(\phi_2^2)^2\right)^3+2\left(-(\phi_2^1)^3+3\phi_2^1 (\phi_2^2)^2\right)
\end{equation}
which has a minimum at $\phi_2=(-\frac{1}{2},\frac{\sqrt{3}}{2})$ as required.

For the flavon $\phi_ 3$, it is easier to work in the non-$\mu$-$\tau$ rotated basis. Defining $\phi_3 = (\phi_3^1,\phi_3^2,\phi_3^3)$ in this basis, $S_4$ invariants can be easily constructed by summing up even powers of $\phi_3^1$, $\phi_3^2$ and $\phi_3^3$ symmetrically. A potential constructed in this way is given below:
\begin{eqnarray}
V(\phi_3)&\propto&\left((\phi_3^1)^8+(\phi_3^2)^8+(\phi_3^3)^8\right) + p \left((\phi_3^1)^6+(\phi_3^2)^6+(\phi_3^3)^6\right)+q \left((\phi_3^1)^4+(\phi_3^2)^4+(\phi_3^3)^4\right)\nonumber\\
&+&r \left((\phi_3^1)^2+(\phi_3^2)^2+(\phi_3^3)^2\right)
+s \left((\phi_3^1)^2 (\phi_3^2)^2+(\phi_3^3)^2 (\phi_3^2)^2+(\phi_3^1)^2 (\phi_3^3)^2\right)
\end{eqnarray}
for real $p$, $q$, $r$ and $s$ under some constraints. This results in $\langle\phi_3\rangle=(1,-\sqrt{2},0)$ in the non-rotated basis, corresponding to $\langle\phi_3\rangle=(1,-1,1)$, in the $\mu$-$\tau$ rotated basis, as required.
Using the extremisation condition of zero first order derivatives applied at the point $\phi_3=(1,-\sqrt{2},0)$, we get the following constraints:
\begin{subequations}
\begin{gather}
r = -60 - 21 p - 6 q,\\
s = 28 + 9 p + 2 q.
\end{gather}
\end{subequations}
To ensure that the extrema points are minima, we need to impose the condition of positive definite Hessian matrix. This gives the following inequalities:
\begin{subequations}
\begin{gather}
4+p>0,\\
18 + 5 p + q>0,\\
208 + 72 p + 9 p^2 - 8 q>0.
\end{gather}
\end{subequations}

The flavons $\phi'_{3e}$, $\phi'_{3\mu}$ and $\phi'_{3\tau}$ belong to $\boldsymbol{3'}$. Let $\phi=(\phi^1, \phi^2, \phi^3)$ transform as a $\boldsymbol{3'}$ in the non-rotated basis of $\boldsymbol{3'}$. Consider a potential 
\begin{eqnarray}\label{eq:cubepot}
V(\phi)&\propto&\left((\phi^1)^8+(\phi^2)^8+(\phi^3)^8\right) + p \left((\phi^1)^6+(\phi^2)^6+(\phi^3)^6\right)+q \left((\phi^1)^4+(\phi^2)^4+(\phi^3)^4\right)\nonumber\\
&+&r \left((\phi^1)^2+(\phi^2)^2+(\phi^3)^2\right)
+s \left((\phi^1)^2 (\phi^2)^2+(\phi^3)^2 (\phi^2)^2+(\phi^1)^2 (\phi^3)^2\right)
\end{eqnarray}
under the constraints
\begin{subequations}\label{eq:cubeconstr1}
\begin{gather}
r = -4 - 3 p - 2 q,\\
s = 0
\end{gather}
\end{subequations}
and 
\begin{subequations}\label{eq:cubeconstr2}
\begin{gather}
-4-3p-2q>0,\\
6+3p+q>0.
\end{gather}
\end{subequations}
The irrep $\boldsymbol{3'}$ denotes the rotational symmetries of a cube. It can be shown that the potential, Eq.~(\ref{eq:cubepot}), with the constraints, Eqs.~(\ref{eq:cubeconstr1},~\ref{eq:cubeconstr2}), has local minima at the centre $(0,0,0)$, vertices ($(1,1,1)$ etc.), edge centres ($(1,1,0)$ etc.), and face centres ($(1,0,0)$ etc.) of the symmetry cube. The potentials for the flavons $\phi'_{3e}$, $\phi'_{3\mu}$ and $\phi'_{3\tau}$ are assumed to be in the same form as Eq.~(\ref{eq:cubepot}). The flavon $\phi'_{3e}$ gets a VEV $(1,0,0)$ which is a face centre. The flavons $\phi'_{3\mu}$ and $\phi'_{3\tau}$ get VEVs $(0,1,1)$ and $(0,-1,1)$ respectively in the non-rotated basis which are edge centres. These edge centres correspond to $(0,\sqrt{2},0)$ and $(0,0,\sqrt{2})$ in the $\mu$-$\tau$ rotated basis. So we have shown that for $\phi'_{3e}$, $\phi'_{3\mu}$ and $\phi'_{3\tau}$, potentials of the form given in Eq.~(\ref{eq:cubepot}) with the constraints, Eqs.~(\ref{eq:cubeconstr1},~\ref{eq:cubeconstr2}), lead to VEVs proportional to $(1,0,0)$, $(0,1,0)$ and $(0,0,1)$ in the $\mu$-$\tau$ rotated basis as originally assumed in Section~3.

\providecommand{\href}[2]{#2}\begingroup\raggedright\endgroup

\end{document}